\documentclass[prl,twocolumn]{revtex4}
\usepackage{dcolumn}
\usepackage{multirow}
\usepackage{graphicx}
\usepackage{amssymb}
\usepackage{bm}
\usepackage{hyperref}
\usepackage{epstopdf}
\usepackage{color}
\usepackage{mathrsfs}
\usepackage{amsmath,amssymb,amsthm}
\usepackage{rotating}
\usepackage{sverb, longtable}
\usepackage{subfigure}

\usepackage[graphicx]{realboxes}
\usepackage{adjustbox}

\begin{document}

\title{Evidence for Cosmological Massive Neutrinos}
\author{Deng Wang}
\email{dengwang@ific.uv.es}
\affiliation{Instituto de F\'{i}sica Corpuscular (CSIC-Universitat de Val\`{e}ncia), E-46980 Paterna, Spain}

\begin{abstract}
A key question in cosmology is whether massive neutrinos exist on cosmic scales. Current cosmological observations have severely compressed the viable range for neutrino masses and even prefer phenomenologically an effective negative mass. This poses a great challenge to the cosmological search for neutrinos. Based on current background and large scale structure data, taking a full redshift and/or scale tomography method, we find one beyond $5\,\sigma$, two $3\,\sigma$ and two $2\,\sigma$ evidences of massive neutrinos, spanning both high and low redshifts, as well as both small and intermediate scales. Interestingly, these five neutrino masses are well consistent within $1\,\sigma$ confidence level, indicating a possible suppression of neutrino mass during the evolution of the universe. Using cosmic microwave background observations to constrain a redshift and scale dependent neutrino mass, we make the first neutrino mass map through the cosmic history and full scales for future high precision search.

\end{abstract}

\maketitle
\section{I. Introduction}
The discovery of neutrino oscillations indicates that neutrinos have mass, which requires new physics beyond the Standard Model. The solar and atmospheric oscillation experiments have measured squared mass differences $\Delta m^2_{21} \approx 7.5\cdot 10^{-5}\,\mathrm{eV}^2$ and $|\Delta m^2_{31}|\approx 2.5\cdot 10^{-3}\,\mathrm{eV}^2$ \cite{KamLAND:2002uet,SNO:2003bmh,Super-Kamiokande:2000ywb,K2K:2002icj,Super-Kamiokande:2003yed}, which means that there exist at least two massive neutrinos and leads to two possible mass hierarchies in the absence of absolute mass scale: (i) normal hierarchy (NH, $\sum m_\nu \gtrsim 0.06$ eV); (ii) inverted hierarchy (IH, $\sum m_\nu \gtrsim 0.10 $ eV) \cite{Hannestad:2016fog,Capozzi:2021fjo,SajjadAthar:2021prg,Esteban:2024eli}. Since these measurements are implemented at local scales, an intriguing and natural problem arises: whether do neutrinos have mass on cosmic scales?  

In theory, neutrinos as one of four main messengers, together with electromagnetic radiation, cosmic rays and gravitational waves, play an important role in the evolution of the universe \cite{Lesgourgues:2006nd,Balantekin:2013gqa,Lattanzi:2017ubx,DeSalas:2018rby,DiValentino:2024xsv}. They interact in the primordial plasma with charged leptons and hadrons via electroweak interactions, until the reaction rates for these processes become so low compared with the typical expansion rate that they decouple and start to propagate freely along the geodesics. They contribute to the background expansion of the universe by the Friedmann equation, while they also affect the large scale structure formation by their masses, abundances and other underlying properties. In light of these physical effects, cosmologists have probed the neutrino mass by using the anisotropies of cosmic microwave background (CMB) \cite{Planck:2018vyg,WMAP:2003elm,Planck:2013pxb}, baryon acoustic oscillations (BAO) \cite{SDSS:2005xqv,2dFGRS:2005yhx,eBOSS:2020yzd,deCarvalho:2017xye,eBOSS:2017cqx,eBOSS:2020gbb}, Type Ia supernovae (SN) \cite{SupernovaSearchTeam:1998fmf,SupernovaCosmologyProject:1998vns}, weak lensing \cite{Heymans:2012gg,Hildebrandt:2016iqg,Planck:2018lbu,DES:2017qwj}, galaxy clustering \cite{DES:2017myr,DES:2021wwk} and Ly$\alpha$ forests \cite{SDSS:2004kqt,McDonald:2006qs,Palanque-Delabrouille:2014jca,Palanque-Delabrouille:2015pga,Yeche:2017upn,BOSS:2017fdr}, which can help constrain the mass sum of three active neutrinos. However, unfortunately, only upper bounds on the neutrino mass are given by various kinds of observations so far \cite{Vagnozzi:2017ovm,Wang:2018ahw,DiValentino:2021hoh}. Specifically, the Planck-2018 CMB temperature, polarization and lensing data gives a $2\,\sigma$ upper limit of $\sum m_\nu < 0.24$ eV \cite{Planck:2018vyg}. The addition of the Sloan Digital Sky Survey (SDSS) BAO data to CMB provides a tighter limit of $\sum m_\nu < 0.12$ eV \cite{Planck:2018vyg}, which puts pressure on the inverted hierarchy. This bound is compatible with constraints from neutrino laboratory experiments which also slightly prefer the normal hierarchy at $2-3\,\sigma$ confidence level \cite{NOvA:2017abs,Super-Kamiokande:2017yvm,Capozzi:2018ubv,deSalas:2017kay}. Interestingly, when combined with CMB temperature, polarization, and lensing observations from the Planck satellite \cite{Planck:2019nip}, as well as lensing observations from the Data Release 6 of the Atacama Cosmology Telescope \cite{ACT:2023kun}, the recent BAO measurements from the first data release of the Dark Energy Spectroscopic Instrument (DESI) \cite{DESI:2024uvr,DESI:2024lzq} produces a very strong $2\,\sigma$ upper bound of $\sum m_\nu < 0.072$ eV \cite{DESI:2024mwx}, which does not support the inverted hierarchy and even poses a strong challenge to the normal hierarchy. Furthermore, considering complementary distance measurements from galaxy clusters \cite{DeFilippis:2005hx} and gamma ray bursts \cite{Demianski:2016zxi}, the most tightest limit of $\sum m_\nu < 0.043$ eV is obtained \cite{Wang:2024hen}, which shows a clear tension between cosmological observations and terrestrial oscillation experiments. Nonetheless, this bound is brought at the price of slightly increasing the present-day expansion rate of the universe. Intriguingly, an effective neutrino mass $\sum \tilde{m}_\nu \simeq -0.16 \pm 0.09$ eV is presented in \cite{Craig:2024tky} by extracting approximately the effect of massive neutrinos on the CMB lensing potential power spectrum. This means that current cosmological data may prefer the phenomenological effects that might be caused by a negative total neutrino mass. It seems that the probability of detecting massive neutrinos becomes gradually lower and lower and even we may never observe them on cosmic scales. However, when implementing constraints on neutrino masses under $\Lambda$CDM, we usually ignore an important fact that the neutrino mass sum is assumed as a constant through the cosmic history and across full scales. Actually, we cannot make sure if neutrino mass remains unchanged over time and space for such a huge, complex and diverse system like the universe. This perspective challenges largely current bounds on neutrino mass and forges newly possible pathways to detect cosmic massive neutrinos. Moreover, driven by curiosity, humanity wants to know the possible distributions of neutrino masses over time and space. Fortunately, even though data precision is limited, current cosmological observations (e.g., CMB \cite{Planck:2018vyg}) allow us to perform such a search. Additionally, according to the inherent developmental laws of the field of cosmology, it will be more and more necessary to search for massive neutrinos in each part of the universe in the future. In this study, we present a full redshift and scale tomography method to achieve this goal by using CMB, large scale structure and background data. We find that neutrinos can have mass on cosmic scales, encompassing both high and low redshifts, as well as both small and intermediate scales.   


\section{II. Data and Method}
We consider the Planck-2018 high-$\ell$ \texttt{plik} temperature (TT) likelihood covering multipoles $30\leqslant\ell\leqslant2508$, polarization (EE), and their cross-correlation (TE) data spanning $30\leqslant\ell\leqslant1996$. We include the low-$\ell$ TT \texttt{Commander} and \texttt{SimAll} EE likelihoods in the range $2\leqslant\ell\leqslant29$ \cite{Planck:2019nip}. Furthermore, we conservatively include the Planck CMB lensing likelihood derived from \texttt{SMICA} maps across $8\leqslant\ell \leqslant400$ \cite{Planck:2018lbu}. We use 12 DESI BAO measurements specified in \cite{DESI:2024mwx} from various galaxy samples spanning the redshift range $0.1  < z < 4.16$ including the BGS sample in $0.1 < z < 0.4$, LRG samples in $0.4 < z < 0.6$ and $0.6 < z < 0.8$, a combined LRG and ELG sample in $0.8 < z < 1.1$, ELG sample in $1.1 < z < 1.6$, a quasar sample in $0.8 < z < 2.1$, and the Ly$\alpha$ forest sample in $1.77 < z < 4.16$ \cite{DESI:2024lzq,DESI:2024uvr}. We take the Pantheon+ SN sample, comprising 1701 light curves of 1550 spectroscopically confirmed SN sourced from eighteen different surveys \cite{Scolnic:2021amr}. We adopt large scale structure observations from the Dark Energy Survey Year 1 (DESY1) including three two point correlation functions, i.e., galaxy clustering, cosmic shear and galaxy-galaxy lensing \cite{DES:2017qwj,DES:2017myr,DES:2017gwu}, which roughly covers the range $z \lesssim 1$. We employ the galaxy power spectra measured at four effective redshfits $z_{eff}=$ 0.22, 0.41, 0.60, and 0.78 from the WiggleZ Dark Energy Survey \cite{Blake:2010xz,Parkinson:2012vd}, which measures 238,000 galaxy redshifts from seven regions of the sky with a total volume of 1 Gpc$^{3}$ in the scale range $[0.01, \, 0.5]$~$h$ Mpc$^{-1}$. Hereafter, we denote CMB, BAO, SN, DESY1 and WiggleZ as ``C'', ``B'', ``S'', ``D'' and ``W'', respectively.

\begin{figure}
	\centering
	\includegraphics[scale=0.5]{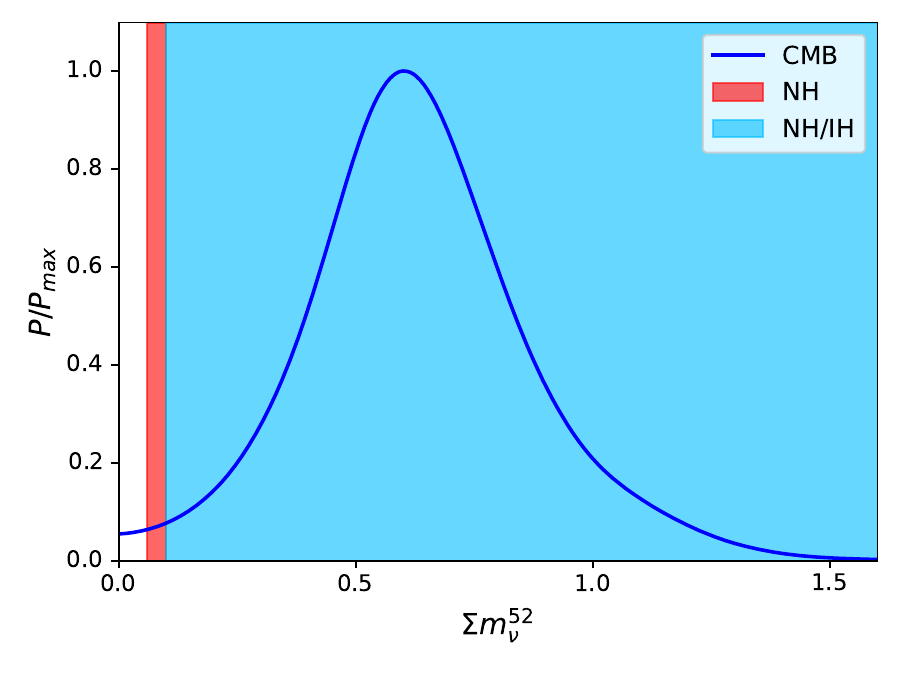}
	\caption{One-dimensional posterior distribution of the parameter $\Sigma m_\nu^{52}$ from CMB observations in $\sum m_\nu(z,\,k)$ model. The red and blue shaded regions denote the NH and NH/IH, respectively.}\label{fig:mnuhighz}
\end{figure}

\begin{figure}
	\centering
	\includegraphics[scale=0.435]{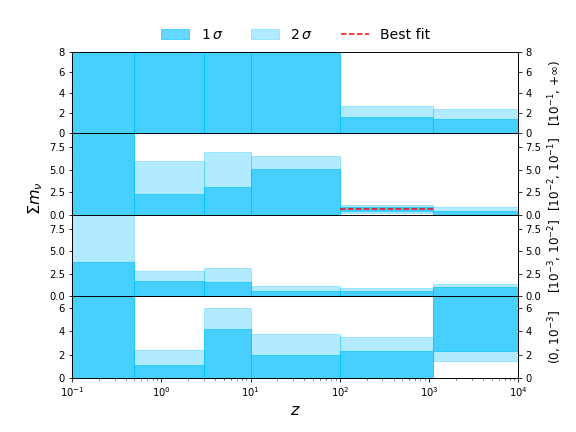}
	\caption{The neutrino mass map over full redshifts and scales from CMB observations in $\sum m_\nu(z,\,k)$ model. The shaded regions are $1\,\sigma$ and $2\,\sigma$ ranges of neutrino mass in each bin. The red line denotes the best fit of $\Sigma m_\nu^{52}$. }\label{fig:mnuzkmap}
\end{figure}

\begin{figure*}
	\centering
	\includegraphics[scale=0.38]{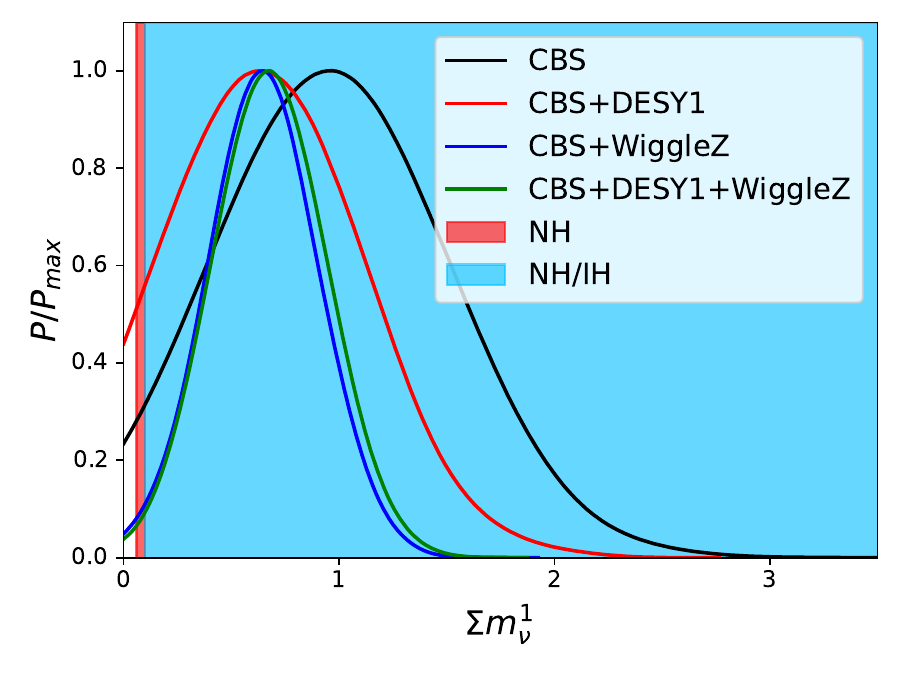}
	\includegraphics[scale=0.38]{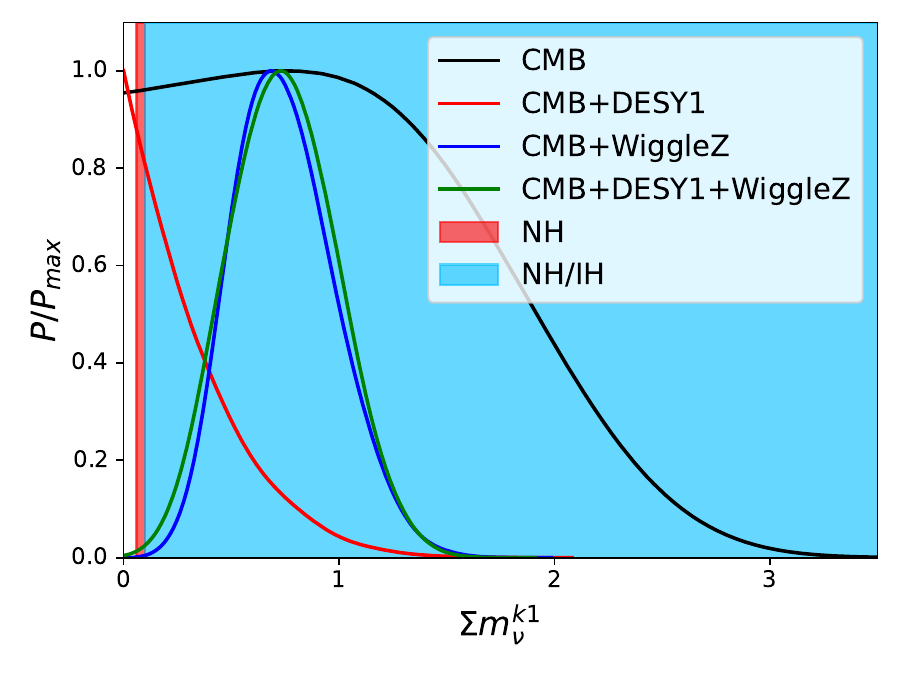}
	\includegraphics[scale=0.38]{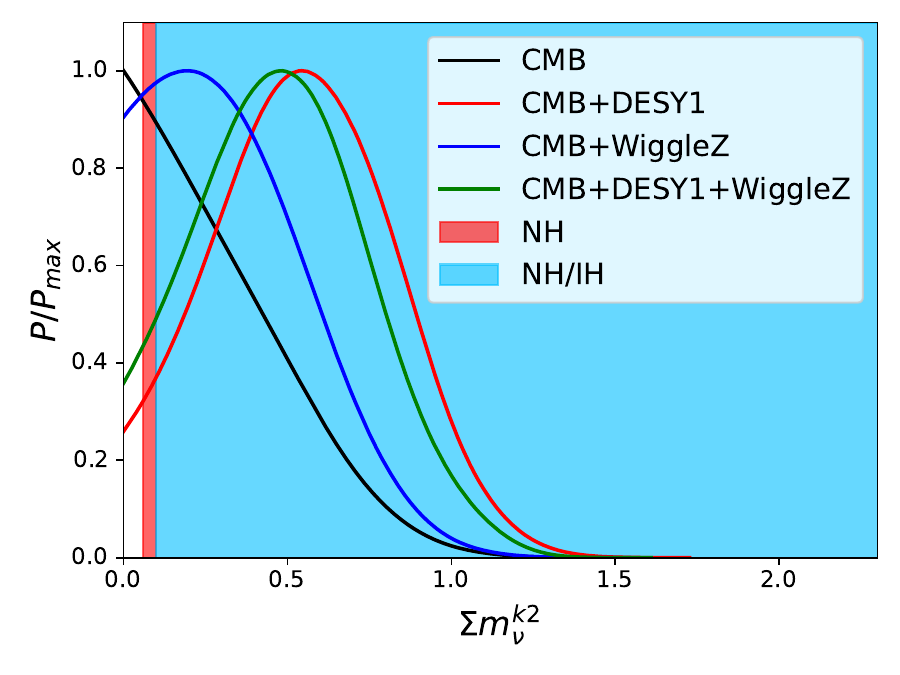}
	\includegraphics[scale=0.38]{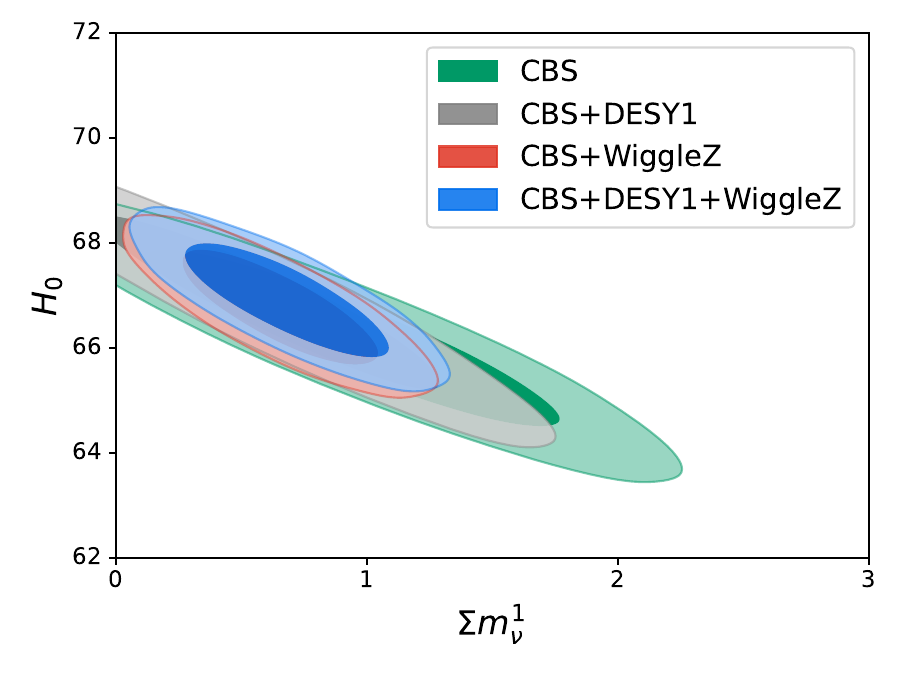}
	\includegraphics[scale=0.38]{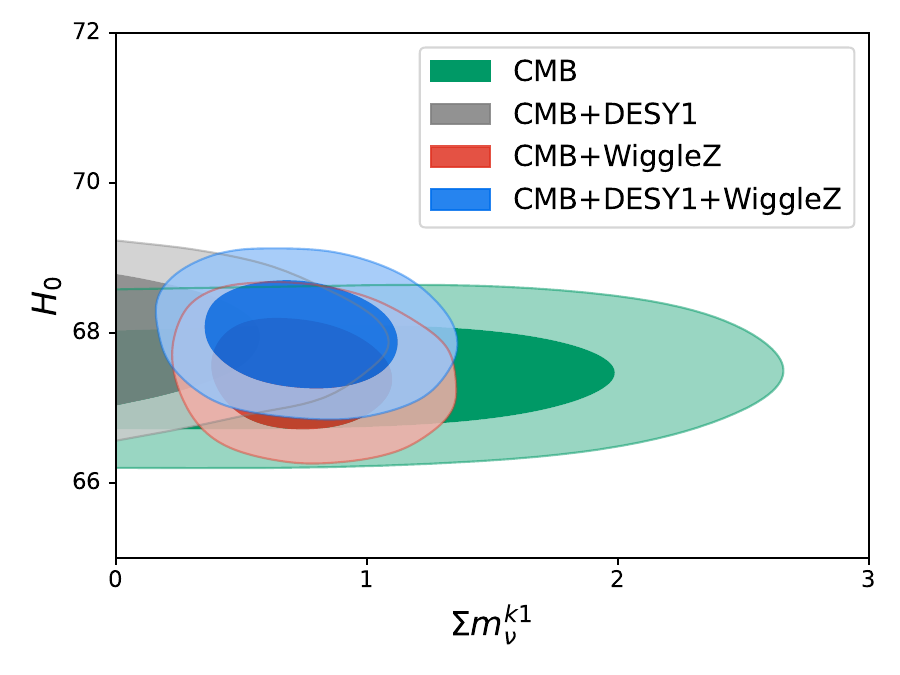}
	\includegraphics[scale=0.38]{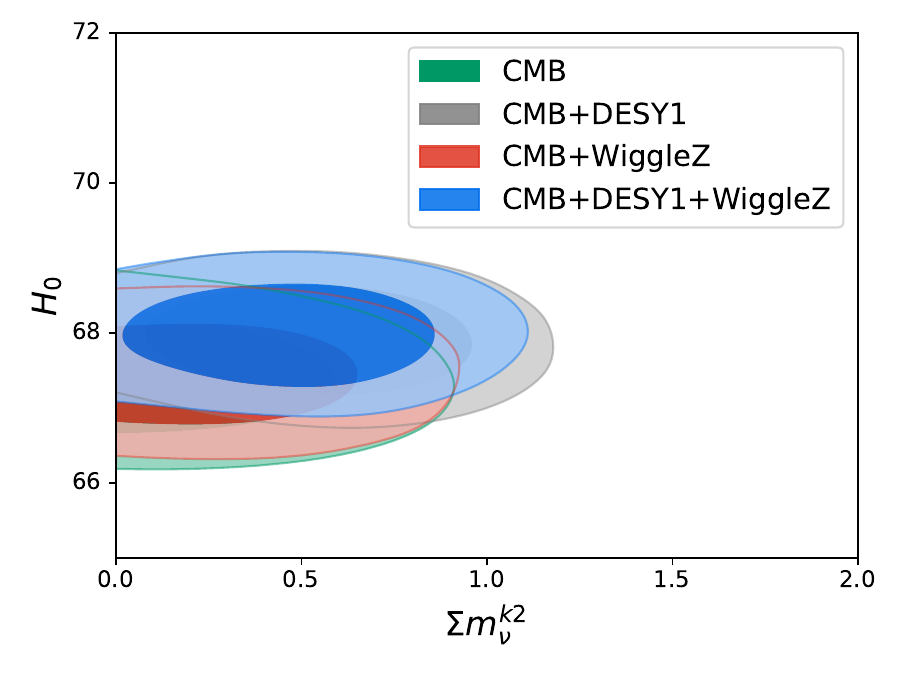}
	\caption{{\it Upper panels}. One-dimensional posterior distributions of the parameters $\Sigma m_\nu^{1}$, $\Sigma m_\nu^{k1}$ and $\Sigma m_\nu^{k2}$ from different datasets in $\sum m_\nu(z)$ and $\sum m_\nu(k)$ models, respectively. {\it Lower panels}. Two-dimensional posterior distributions of the parameter pairs ($\Sigma m_\nu^{1}$, $H_0$), ($\Sigma m_\nu^{k1}$, $H_0$) and ($\Sigma m_\nu^{k2}$, $H_0$) from different datasets in $\sum m_\nu(z)$ and $\sum m_\nu(k)$ models, respectively.}\label{fig:mnulowz}	
\end{figure*}

\begin{table*}[!t]
	\renewcommand\arraystretch{1.5}
	\caption{The $1\,\sigma$ (68\%) errors and mean values of the parameters $\Sigma m_\nu^{52}$, $\Sigma m_\nu^{1}$, $\Sigma m_\nu^{k1}$ and $\Sigma m_\nu^{k2}$ from different datasets in  $\sum m_\nu(z,\, k)$, $\sum m_\nu(z)$ and $\sum m_\nu(k)$ models, respectively.}
	\setlength{\tabcolsep}{2.8mm}{
		\begin{tabular} { l |c| c |c| c| c }
			\hline
			\hline
			
			Model              & $\sum m_\nu(z,\,k)$  & \multicolumn{2}{c}{$\sum m_\nu(z)$ }     &   \multicolumn{2}{|c}{$\sum m_\nu(k)$ }                              \\
			\hline
			Data              &  C      & CBS  & CBSW   & CW   & CD   \\
			\hline
			Parameter         & $\Sigma m_\nu^{52}=0.63^{+0.20}_{-0.24}$     & $\Sigma m_\nu^{1}=1.01^{+0.47}_{-0.58}$      & $\Sigma m_\nu^{1}=0.65\pm 0.25$          &$\Sigma m_\nu^{k1}=0.75^{+0.20}_{-0.27}$    & $\Sigma m_\nu^{k2}=0.55\pm 0.27$      \\
			
			\hline
			\hline
	\end{tabular}}
	\label{t1}
\end{table*}

To characterize the properties of massive neutrinos over full redshifts and scales, we adopt the following three scenarios: (i) redshift dependent $\sum m_\nu(z)$; (ii) scale dependent $\sum m_\nu(k)$; (iii) redshift and scale dependent $\sum m_\nu(z,\,k)$. For the former case, we divide the whole cosmic history into six redshift bins: [0, 1], [1, 3], [3, 10], [10, 100], [100, 1100] and $[1100, +\infty)$. The first bin covers the range where dark energy starts to play a role during the evolution of the universe. The second bin spans roughly the BAO interval. The third bin includes most of the remaining information embedded in CMB lensing \cite{Carbone:2007yy}. The fourth bin describes the era starting from the time when massive neutrinos become non-relativistic to the formation of the first objects. The fifth bin depicts a long period during the dark ages era since CMB photons decouple. The last bin integrates all the information from pre-recombination physics. Notice that, to capture completely the neutrino mass information when dark energy exists, we take [0, 1] instead of [0, 0.5] used in \cite{Lorenz:2021alz}. We use $\Sigma m_\nu^{i}$ with $i=1, ..., 6$ in each $z$ bin. For the middle case, we divide the full scales into four bins: $[10^{-1}, +\infty)$, $[10^{-2}, 10^{-1}]$, $[10^{-3}, 10^{-2}]$ and $[0, 10^{-3}]$ h\,Mpc$^{-1}$. We accordingly assign $\Sigma m_\nu^{ki}$ with $i=1, ..., 4$ to each $k$ bin. For the latter case, we divide full redshifts and scales into six $z$ bins: [0, 0.5], [1, 3], [3, 10], [10, 100], [100, 1100] and $[1100, +\infty)$ and each one has four corresponding $k$ bins: $[10^{-1}, +\infty)$, $[10^{-2}, 10^{-1}]$, $[10^{-3}, 10^{-2}]$ and $[0, 10^{-3}]$ h\,Mpc$^{-1}$. We use $\Sigma m_\nu^{ij}$ with $i=1, ..., 6$ and $j=1, ..., 4$ in each $(z,\,k)$ bin.

In order to compute the background evolution of the universe and theoretical power spectra, we take the Boltzmann code \texttt{CAMB} \cite{Lewis:1999bs}. To perform the Bayesian analysis, we use the Monte Carlo Markov Chain (MCMC) method to infer the posterior probability distributions of model parameters via the publicly available package \texttt{CosmoMC} \cite{Lewis:2002ah,Lewis:2013hha}. We assess the convergence of MCMC chains using the Gelman-Rubin criterion $R-1\lesssim 0.05$ \cite{Gelman:1992zz} and analyze them using the package \texttt{Getdist} \cite{Lewis:2019xzd}.

We take the following uniform priors for the model parameters: the baryon fraction $\Omega_bh^2 \in [0.005, 0.1]$, cold dark matter fraction $\Omega_ch^2 \in [0.001, 0.99]$, acoustic angular scale at the recombination epoch $100\theta_{MC} \in [0.5, 10]$, scalar spectral index $n_s \in [0.8, 1.2]$, amplitude of the primordial scalar power spectrum $\ln(10^{10}A_s) \in [2, 4]$, optical depth $\tau \in [0.01, 0.8]$, and mass sum of three active neutrinos $\Sigma m_\nu \in [0, 5]$ eV. 
For $\sum m_\nu(z)$, we use $\Sigma m_\nu^{i} \in [0, 30]$ eV with $i=1, ..., 6$ in each $z$ bin. 
For $\sum m_\nu(k)$, we adopt $\Sigma m_\nu^{ki} \in [0, 30]$ eV with $i=1, ..., 4$ in each $k$ bin. 
For $\sum m_\nu(z,\,k)$, we employ $\Sigma m_\nu^{ij} \in [0, 30]$ eV with $i=1, ..., 6$ and $j=1, ..., 4$ in each bin. We take the degenerate neutrino hierarchy throughout this study.







\section{III. High-$z$ and low-$z$ evidences}
To study the neutrino mass distributions over redshifts and scales, we constrain the 24-bin $\sum m_\nu(z,\,k)$ model with CMB data that serves as a unique probe of high-$z$ universe. Interestingly, we find $\Sigma m_\nu^{52}=0.63^{+0.20 (1\sigma)+0.52(2\sigma)}_{-0.24(1\sigma)-0.46(2\sigma)}$ eV, indicating a $\sim3\,\sigma$ evidence of massive neutrinos when $z \in [100, 1100]$ and $k \in [10^{-2}, 10^{-1}]$ h\,Mpc$^{-1}$ (see Fig.\ref{fig:mnuhighz} and Tab.\ref{t1}). The corresponding neutrino mass map through cosmic history and across full scales is presented in Fig.\ref{fig:mnuzkmap}. Overall, CMB gives tighter constraints at intermediate scales and high redshifts. When $k \in [10^{-3}, 10^{-2}]$ h\,Mpc$^{-1}$, CMB prefers stronger constraints than other $k$ bins spanning the entire history of the universe. The tightest $2\,\sigma$ upper bound is $\Sigma m_\nu^{62}<0.84$ eV from the pre-recombination era. Furthermore, confronting the 6-bin $\sum m_\nu(z)$ model with different datasets, we find CBS gives $\Sigma m_\nu^{1}=1.01^{+0.47(1\sigma)+0.89(2\sigma)}_{-0.58(1\sigma)-1.00(2\sigma)}$ eV (see Fig.\ref{fig:mnulowz}), implying a $2\,\sigma$ evidence of massive neutrinos in $z \in [0, 1]$. Especially, DESI BAO data dominates the constraints in $z \in [0, 100]$ and the tightest $2\,\sigma$ bound is $\Sigma m_\nu^{4}<0.24$ eV, which also contributes most to recent DESI's global bound of $\Sigma m_\nu<0.072$ eV \cite{DESI:2024mwx}. Attractively, the addition of WiggleZ galaxy power spectra to CBS gives $\Sigma m_\nu^{1}=0.65^{+0.25(1\sigma)+0.51(2\sigma)+0.67(3\sigma)}_{-0.25(1\sigma)-0.49(2\sigma)-0.60(3\sigma)}$ eV in $z \in [0, 1]$, which is a $3\,\sigma$ detection of neutrino mass. Although preferring a lower value, it is consistent with CBS constraint within $1\,\sigma$ level. Interestingly, we find that the $\sum m_\nu$-$H_0$ anti-correlation sources from the degeneracy between dark energy and massive neutrinos when $z \lesssim 1$.
Since WiggleZ and DESY1 are low-$z$ surveys ($z \lesssim 1$), we regard their constraints on $\sum m_\nu(k)$ as low-$z$ ones. For the 4-bin $\sum m_\nu(k)$ model, CW gives $\Sigma m_\nu^{k1}=0.75^{+0.20(1\sigma)+0.48(2\sigma)+0.67(3\sigma)}_{-0.27(1\sigma)-0.44(2\sigma)-0.52(3\sigma)}$ eV, which is a beyond $5\,\sigma$ evidence of massive neutrinos at small scales in $k \in [10^{-1}, +\infty)$ h\,Mpc$^{-1}$, while CD gives $\Sigma m_\nu^{k2}=0.55^{+0.27(1\sigma)+0.45(2\sigma)}_{-0.27(1\sigma)-0.54(2\sigma)}$ eV implying a $2\,\sigma$ evidence at intermediate scales in $k \in [10^{-2}, 10^{-1}]$ h\,Mpc$^{-1}$. So far, all the evidences occur at small scales as predicted by theory. Note that our results cannot distinguish NH from IH. 
More details about constraints are shown in the supplementary material (SM). Some further analyses are presented in \cite{Wang:2025ker}.

\section{IV. Big fish eat little fish (BFELF)}
To explain how a global constraint emerges from various redshift and scale bins, we propose the ``BFELF effect'', which states that strong constraining power in a bin swallows weak constraining power in the other bin and becomes stronger in a wider bin. A simple example in $\sum m_\nu(z,\,k)$ model is that $\Sigma m_\nu^{62}<0.84$ eV firstly swallows other three $(z,\,k)$ bins in $k \in [10^{-2}, 10^{-1}]$ h\,Mpc$^{-1}$, then swallows other five $z$ bins and finally becomes the strongest (or global) bound $\Sigma m_\nu<0.24$ eV. A subtle example is why CD gives no signal of $\Sigma m_\nu^{k1}$ but CW provides a beyond $5\,\sigma$ evidence. In $k \in [10^{-1}, +\infty)$ h\,Mpc$^{-1}$, the constraining power of CD mainly sources from $[0.1, 0.2]$ h\,Mpc$^{-1}$, where CD gives a consistent constraint with CW at $1\,\sigma$ level but only has an upper bound. In $[0.2, +\infty)$ h\,Mpc$^{-1}$, CD gives a looser constraint than CW and still provides no signal. Interestingly, after merging these two subbins into $[10^{-1}, +\infty)$ h\,Mpc$^{-1}$, CD gives a stronger bound due to BFELF effect. However, CW give compatible signals in these two subbins, and consequently leads to seemingly inconsistent constraints with CD (see Fig.\ref{fig:mnulowz}). Actually, there is no obvious inconsistency here. The BFELF effect behaves like an emergent one and could be applied in any problem in any field that involves statistical constraints. 


\section{V. Neutrino mass suppression}
Based on the above constraints on $\sum m_\nu(z)$, $\sum m_\nu(k)$ and $\sum m_\nu(z,\,k)$ models using various datasets, we propose four possible evolutionary mechanisms of neutrino mass. Since CMB can independently probe physics over full redshifts and scales, the first one is a constant $\sum m_\nu$ still allowed by CMB constraint on $\sum m_\nu(z,\,k)$. In $z$ direction, although DESI BAO has largely compressed the neutrino mass in $[0, 100]$, we still observe a signal of $\Sigma m_\nu^{1}$ from CBS or CBSW in $[0, 1]$. Subsequently, we also find a signal of $\Sigma m_\nu^{52}$ from CMB in $[100, 1100]$. This means that there may exist a mass suppression in $[1, 100]$, which could be interpreted by neutrino decay \cite{Escudero:2019gfk,Chacko:2019nej,Chacko:2020hmh,Escudero:2020ped}, long-range neutrino forces \cite{Esteban:2021ozz}, neutrino cooling and heating \cite{Lorenz:2021alz} and other novel models \cite{Dvali:2016uhn}. Hence, the second scenario is freedom-suppression-freedom $\sum m_\nu(z)$. However, we cannot ensure if two masses here are same. But, current data supports at least they are same within $1\,\sigma$ level. The third possibility is freedom-suppression, i.e., neutrino mass is suppressed by some unknown mechanism at high $z$ and only appears at low $z$ (e.g., $z \in [0, 1]$). The final one is suppression-freedom, i.e., neutrinos have mass at high $z$ and is always suppressed until today. Making full use of integrated information (see Tab.\ref{t1}) from multiple probes in three models, we prefer the second scenario. The schematic plots of four mechanisms are shown in SM. Note that we cannot determine if there is a suppression of neutrino masses across scales, since large scale constraints are very poor and we only find evidences of massive neutrinos at small scales. 

\section{VI. Discussions and conclusions}
The $\sim3\,\sigma$ detection of $\Sigma m_\nu^{52}$ is the first high-$z$ evidence of massive neutrinos that could be related to the anomalous amplitude of integrated Sachs-Wolfe effect during dark ages \cite{Wang:2024kpu}. Very interestingly, five evidences in Tab.\ref{t1} are consistent with each other at $1\,\sigma$ level and prefers roughly a value of $\sim 0.6$ eV. This means that massive neutrinos may truly exist in a non-common way that cannot be derived from the traditional single parameter constraint. Recently, the finding that the Weyl potential measured by the first three years of DES observations is lower than the $\Lambda$CDM prediction at low redshifts, reveals a beyond $2\,\sigma$ suppression of structure formation \cite{Tutusaus:2023aux}, which is well consistent with our low-$z$ evidences of massive neutrinos in $\sum m_\nu(z)$ and $\sum m_\nu(k)$ models.

For $\sum m_\nu(k)$ in $k \in [10^{-1}, +\infty)$ h\,Mpc$^{-1}$, the $2\,\sigma$ lower limit of $\Sigma m_\nu^{k1}>0.31$ eV from CW helps rule out the effective Majorana mass scenario when combined with 90\% upper limit of 36-156 meV from the KamLAND-Zen's neutrinoless double beta decay search \cite{KamLAND-Zen:2022tow}. Overall, we find evidences of massive neutrinos at both high and low redshifts, while detecting them at both small and intermediate scales where they suppress the structure formation. Attractively, we not only find signals (e.g., $\Sigma m_\nu^{52}$, $\Sigma m_\nu^{k1}$ and $\Sigma m_\nu^{k2}$) at the perturbation level using large scale structure datasets such as WiggleZ and DESY1 (i.e., CW and CD), but also give evidences at the background level using BAO and SN (i.e., CBS). This shows the diversity of our analyses and, more importantly, verifies the stability of our conclusions. It is noteworthy that our results cannot clearly help determine which hierarchy is correct. Even if combined with KATRIN's 90\% upper limit of $0.8$ eV \cite{KATRIN:2021uub}, our constraints cannot distinguish NH from IH. The ongoing and future observations \cite{Archidiacono:2016lnv,Brinckmann:2018owf} will continuously compressing data precision and neutrino mass parameter space and help go for a further step to explore the neutrino hierarchy.  


In light of the fact that DESI BAO data puts a strong pressure on positive neutrino masses \cite{DESI:2024mwx} and current observations prefer phenomenologically $\Sigma m_\nu<0$ eV \cite{Craig:2024tky}, our tomographic full redshifts and scales analyses demonstrate that neutrinos can have masses on cosmic scales and consequently, to a large extent, solve the crisis of phenomenologically negative mass. Furthermore, by proposing the BFELF effect, we clarify the status of current neutrino mass constraints, and explain how one obtains the global tight constraint on $\Sigma m_\nu$ through cosmic history and full scales. Driven by the future demand for high-precision data applications, we propose a simple $\sum m_\nu(z,\,k)$ framework of depicting the neutrino mass distribution of the universe so as to map the neutrino sky better.  

\section*{Acknowledgements}
DW is supported by the CDEIGENT fellowship of Consejo Superior de Investigaciones Científicas (CSIC).




\clearpage

\appendix

\onecolumngrid
\section{\large Appendix}
\twocolumngrid

In this appendix, first of all, we present the constraints on $\sum m_\nu$ from different datasets. Then, we show the constraints on our tomographic neutrino mass models including $\sum m_\nu(z,\,k)$, $\sum m_\nu(k)$ and $\sum m_\nu(z)$. Furthermore, based on the integrated information from our constraints, we depict four possible evolutionary mechanisms of neutrino masses through the cosmic history. Finally, we demonstrate that our evidences of massive neutrinos are not induced by the so-called $S_8$ (or $\sigma_8$) tension \cite{DiValentino:2020vvd1}, because both DESY1 and WiggleZ independently prefer a non-zero neutrino mass at small scales in the absence of CMB data. Hereafter same as the main text, we refer to CMB, BAO, SN, DESY1 and WiggleZ as ``C'', ``B'', ``S'', ``D'' and ``W'', respectively.

\subsection{A. Global constraints on $\sum m_\nu$} 

\begin{figure*}[htbp]
	\centering
	\includegraphics[scale=0.5]{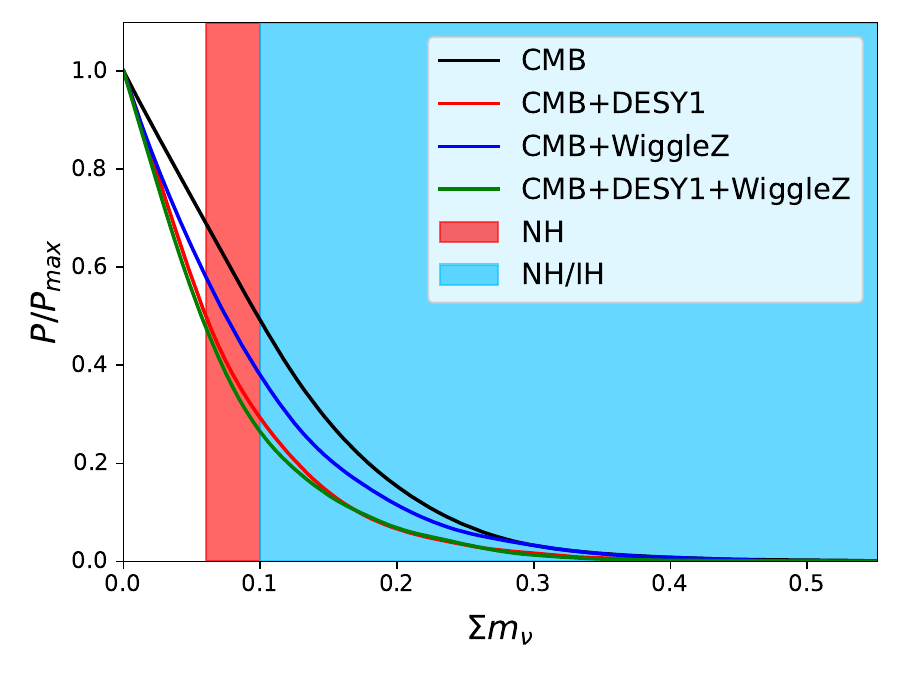}
	\includegraphics[scale=0.5]{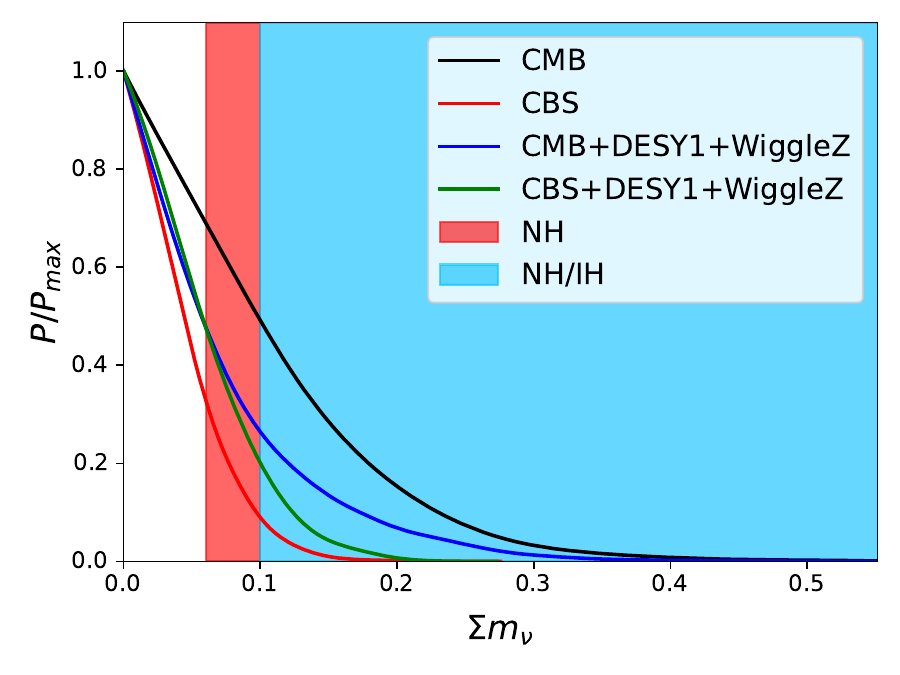}
	\includegraphics[scale=0.5]{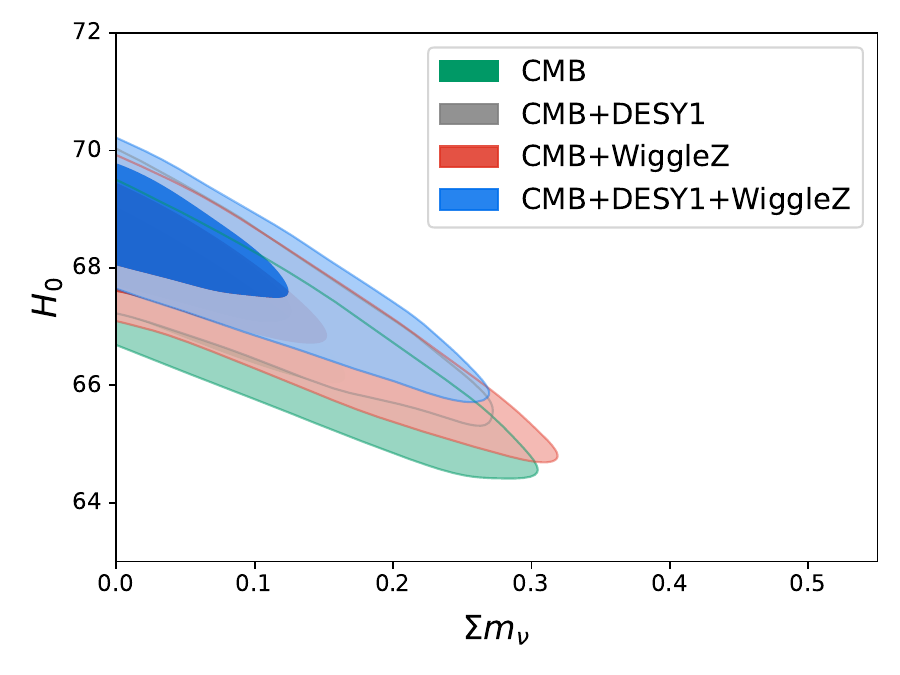}
	\includegraphics[scale=0.5]{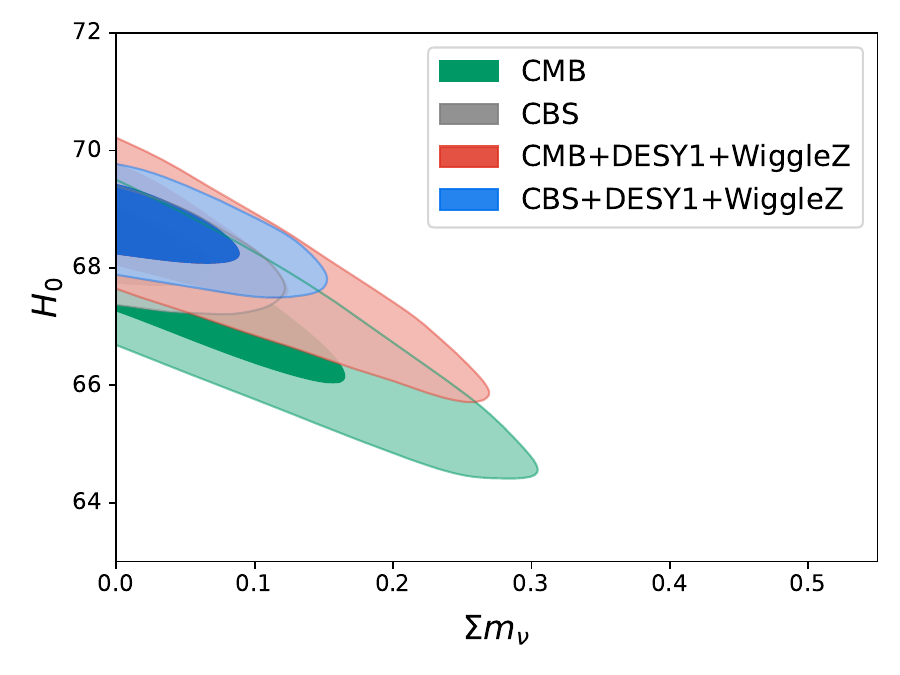}
	\caption{{\it Upper panels}. One-dimensional posterior distributions of the parameter $\Sigma m_\nu$ from different datasets in the $\Lambda$CDM model. The red and blue shaded regions denote the NH and NH/IH, respectively. {\it Lower panels}. Two-dimensional posterior distributions of the parameter pair ($\Sigma m_\nu$, $H_0$) from different data combinations in the $\Lambda$CDM model.}\label{fig:mnu}
\end{figure*}

\begin{figure*}[htbp]
	\centering
	\includegraphics[scale=0.145]{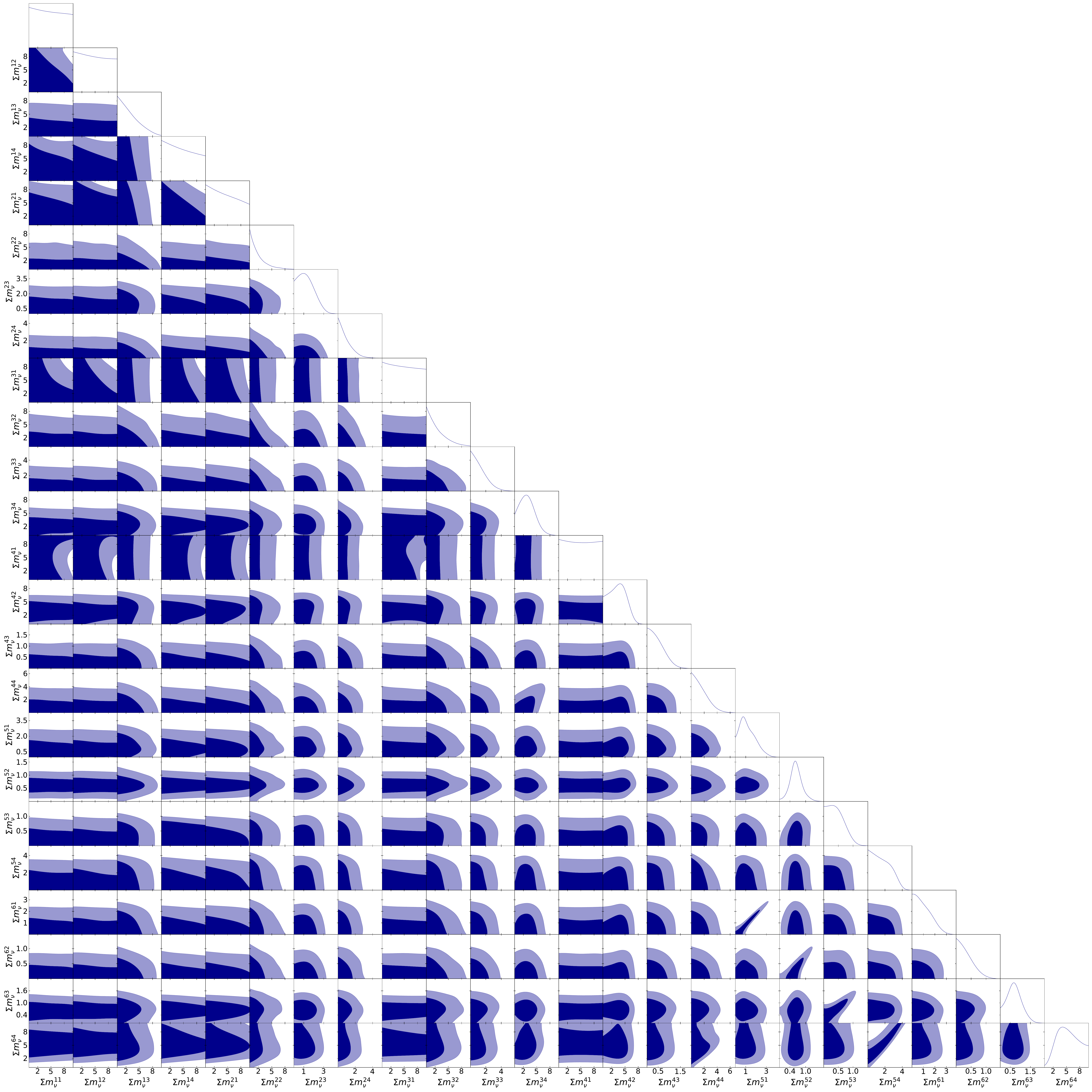}
	\caption{One-dimensional and two-dimensional posterior distributions of the neutrino mass parameters from CMB observations in the $\sum m_\nu(z,\,k)$ model.}\label{fig:mnuzk}
\end{figure*}

\begin{table*}[!t]
	\renewcommand\arraystretch{1.6}
	\caption{The $2\,\sigma$ ($95\%$) upper or lower limits of the neutrino mass parameters from CMB observations in the $\sum m_\nu(z,\,k)$ model. Note that we quote the mean value and $1\,\sigma$ ($68\%$) uncertainty for $\sum m_\nu^{52}$ and the symbols ``$\heartsuit$'' denote the parameters that cannot be constrained by data.}  
	\setlength{\tabcolsep}{5mm}{
		\begin{tabular} { l | c}
			\hline
			\hline
			Parameter              &  Limit  \\
			\hline
			
			{\boldmath$\Sigma m_\nu^{11}$} & $\heartsuit$                         \\
			
			{\boldmath$\Sigma m_\nu^{12}$} & $\heartsuit$                         \\
			
			{\boldmath$\Sigma m_\nu^{13}$} & $< 7.30                    $\\
			
			{\boldmath$\Sigma m_\nu^{14}$} & $\heartsuit$                         \\
			
			{\boldmath$\Sigma m_\nu^{21}$} & $\heartsuit$                         \\
			
			{\boldmath$\Sigma m_\nu^{22}$} & $< 5.90                    $\\
			
			{\boldmath$\Sigma m_\nu^{23}$} & $< 2.69                    $\\
			
			{\boldmath$\Sigma m_\nu^{24}$} & $< 2.40                    $\\
			
			{\boldmath$\Sigma m_\nu^{31}$} & $\heartsuit$                         \\
			
			{\boldmath$\Sigma m_\nu^{32}$} & $< 6.89                    $\\
			
			{\boldmath$\Sigma m_\nu^{33}$} & $< 3.12                    $\\
			
			{\boldmath$\Sigma m_\nu^{34}$} & $< 5.88                    $\\
			
			{\boldmath$\Sigma m_\nu^{41}$} & $\heartsuit$                         \\
			
			{\boldmath$\Sigma m_\nu^{42}$} & $< 6.31                    $\\
			
			{\boldmath$\Sigma m_\nu^{43}$} & $< 1.10                    $\\
			
			{\boldmath$\Sigma m_\nu^{44}$} & $< 3.75                    $\\
			
			{\boldmath$\Sigma m_\nu^{51}$} & $< 2.59                    $\\
			
			{\boldmath$\Sigma m_\nu^{52}$} & $0.63^{+0.20}_{-0.24}      $\\
			
			{\boldmath$\Sigma m_\nu^{53}$} & $< 0.912                   $\\
			
			{\boldmath$\Sigma m_\nu^{54}$} & $< 3.47                    $\\
			
			{\boldmath$\Sigma m_\nu^{61}$} & $< 2.34                    $\\
			
			{\boldmath$\Sigma m_\nu^{62}$} & $< 0.84                   $\\
			
			{\boldmath$\Sigma m_\nu^{63}$} & $< 1.33                    $\\
			
			{\boldmath$\Sigma m_\nu^{64}$} & $> 1.76                    $\\
			
			\hline
			\hline
		\end{tabular}
		\label{tab:mnuzk}}
\end{table*}

\begin{figure*}[htbp]
	\centering
	\includegraphics[scale=0.6]{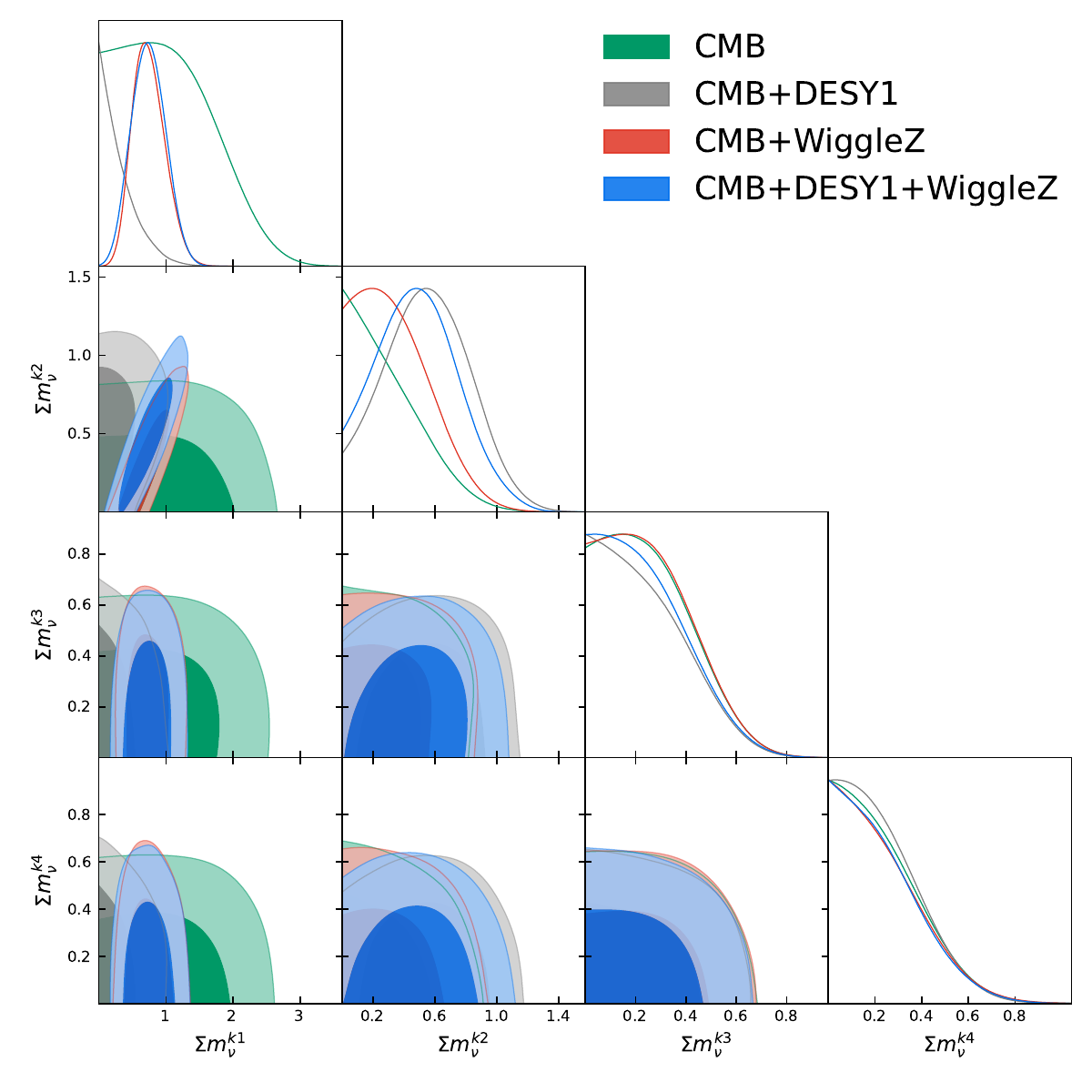}
	\caption{One-dimensional and two-dimensional posterior distributions of the neutrino mass parameters from different datasets in the $\sum m_\nu(k)$ model.}\label{fig:mnuk}
\end{figure*}

\begin{table*}[!t]
	\renewcommand\arraystretch{1.6}
	\caption{The mean values and $1\,\sigma$ ($68\%$) uncertainties of the neutrino mass parameters from different datasets in the $\sum m_\nu(k)$ model. Note that we quote $2\,\sigma$ ($95\%$) upper limits of neutrino masses for parameters without $1\,\sigma$ uncertainties.}
	\setlength{\tabcolsep}{7mm}{
		\begin{tabular} { l |c| c |c| c }
			\hline
			\hline
			Parameter              &  CMB & CMB+DESY1 & CMB+WiggleZ & CMB+DESY1+WiggleZ    \\
			\hline
			
			$\Sigma m_\nu^{k1}         $ & $< 2.22                    $ & $< 0.841                   $ & $0.75^{+0.20}_{-0.27}      $ & $0.74\pm 0.25              $\\
			
			$\Sigma m_\nu^{k2}         $ & $< 0.737                   $ & $0.55\pm 0.27       $ & $< 0.768                   $ & $< 0.930                   $\\
			
			$\Sigma m_\nu^{k3}         $ & $< 0.564                   $ & $< 0.553                   $ & $< 0.562                   $ & $< 0.552                   $\\
			
			$\Sigma m_\nu^{k4}         $ & $< 0.553                   $ & $< 0.540                   $ & $< 0.559                   $ & $< 0.552                   $\\
			\hline
			\hline
		\end{tabular}
		\label{tab:mnuk}}
\end{table*}

\begin{figure*}[htbp]
	\centering
	\includegraphics[scale=0.5]{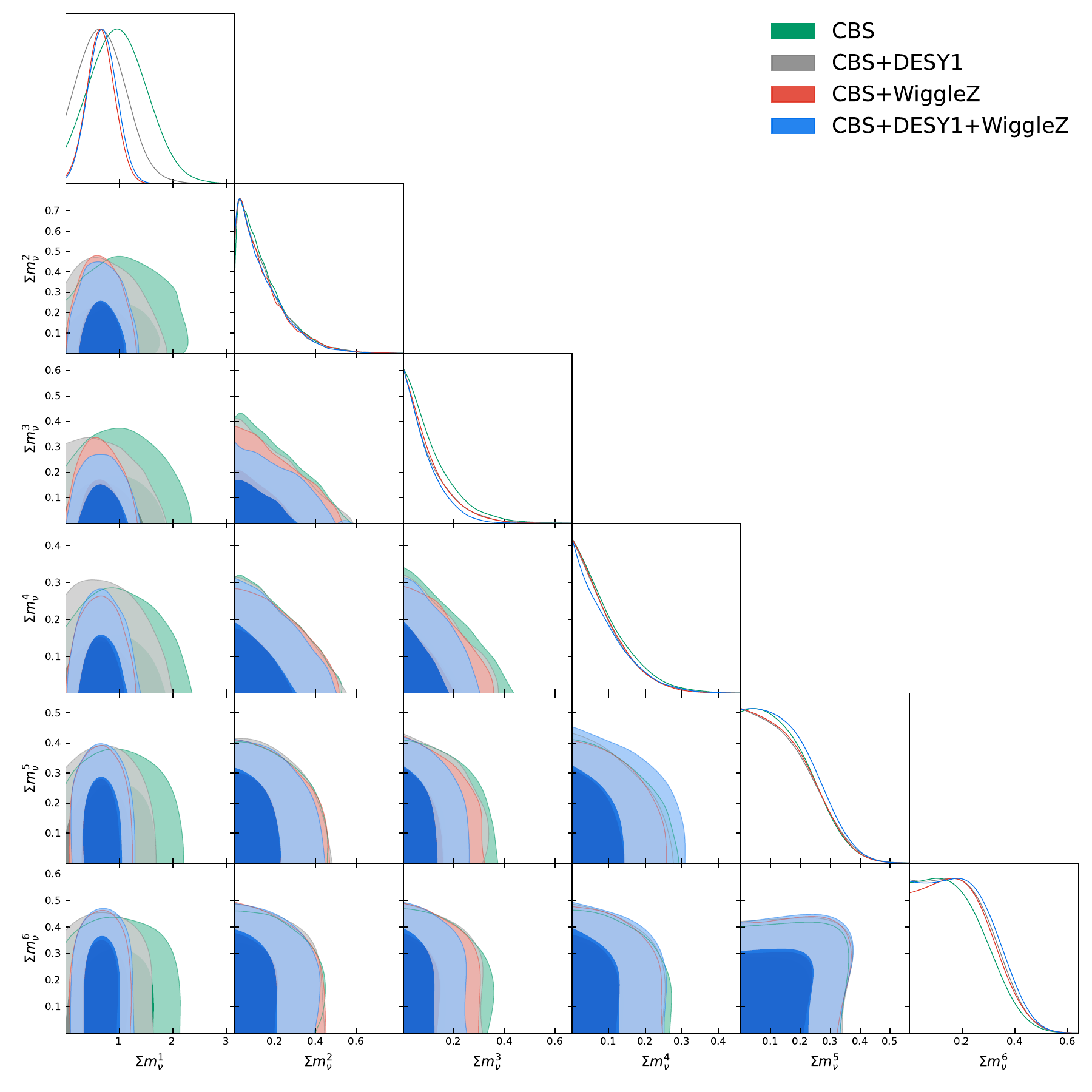}
	\caption{One-dimensional and two-dimensional posterior distributions of the neutrino mass parameters from different datasets in the $\sum m_\nu(z)$ model.}\label{fig:mnuz}
\end{figure*}

\begin{table*}[!t]
	\renewcommand\arraystretch{1.6}
	\caption{The mean values and $1\,\sigma$ ($68\%$) uncertainties of the neutrino mass parameters from different datasets in the $\sum m_\nu(k)$ model. Note that we quote $2\,\sigma$ ($95\%$) upper limits of neutrino masses for parameters without $1\,\sigma$ uncertainties.}
	\setlength{\tabcolsep}{7mm}{
		\begin{tabular} { l |c| c |c| c }
			\hline
			\hline
			Parameter              &  CBS & CBS+DESY1 & CBS+WiggleZ & CBS+DESY1+WiggleZ    \\
			\hline
			
			$\Sigma m_\nu^1            $ & $1.01^{+0.47}_{-0.58}      $ & $<1.44      $ & $0.65\pm 0.25              $ & $0.68\pm 0.26              $\\
			
			$\Sigma m_\nu^2            $ & $< 0.384                   $ & $< 0.385                   $ & $< 0.390                   $ & $< 0.368                   $\\
			
			$\Sigma m_\nu^3            $ & $< 0.306                   $ & $< 0.280                   $ & $< 0.267                   $ & $< 0.221                   $\\
			
			$\Sigma m_\nu^4            $ & $< 0.234                   $ & $< 0.222                   $ & $< 0.215                   $ & $< 0.226                   $\\
			
			$\Sigma m_\nu^5            $ & $< 0.323                   $ & $< 0.332                   $ & $< 0.325                   $ & $< 0.330                   $\\
			
			$\Sigma m_\nu^6            $ & $< 0.376                   $ & $< 0.391                   $ & $< 0.391                   $ & $< 0.401                   $\\
			
			\hline
			\hline
		\end{tabular}
		\label{tab:mnuz}}
\end{table*}

\begin{figure*}[htbp]
	\centering
	\includegraphics[scale=0.55]{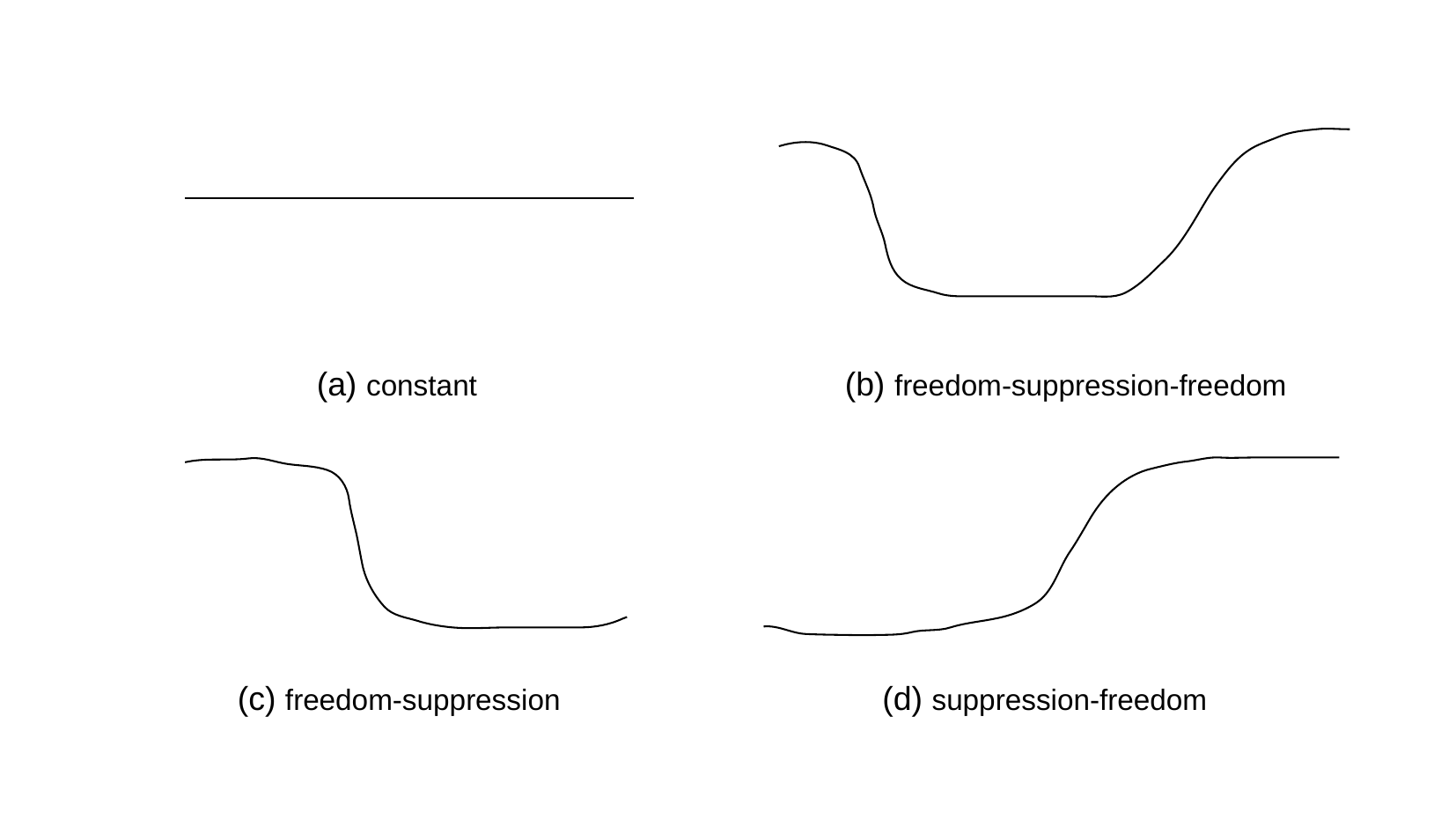}
	\caption{The schematic plots of four evolutionary mechanisms of neutrino masses through the cosmic history.}\label{fig:mechanisms}
\end{figure*}

\begin{figure*}[htbp]
	\centering
	\includegraphics[scale=0.6]{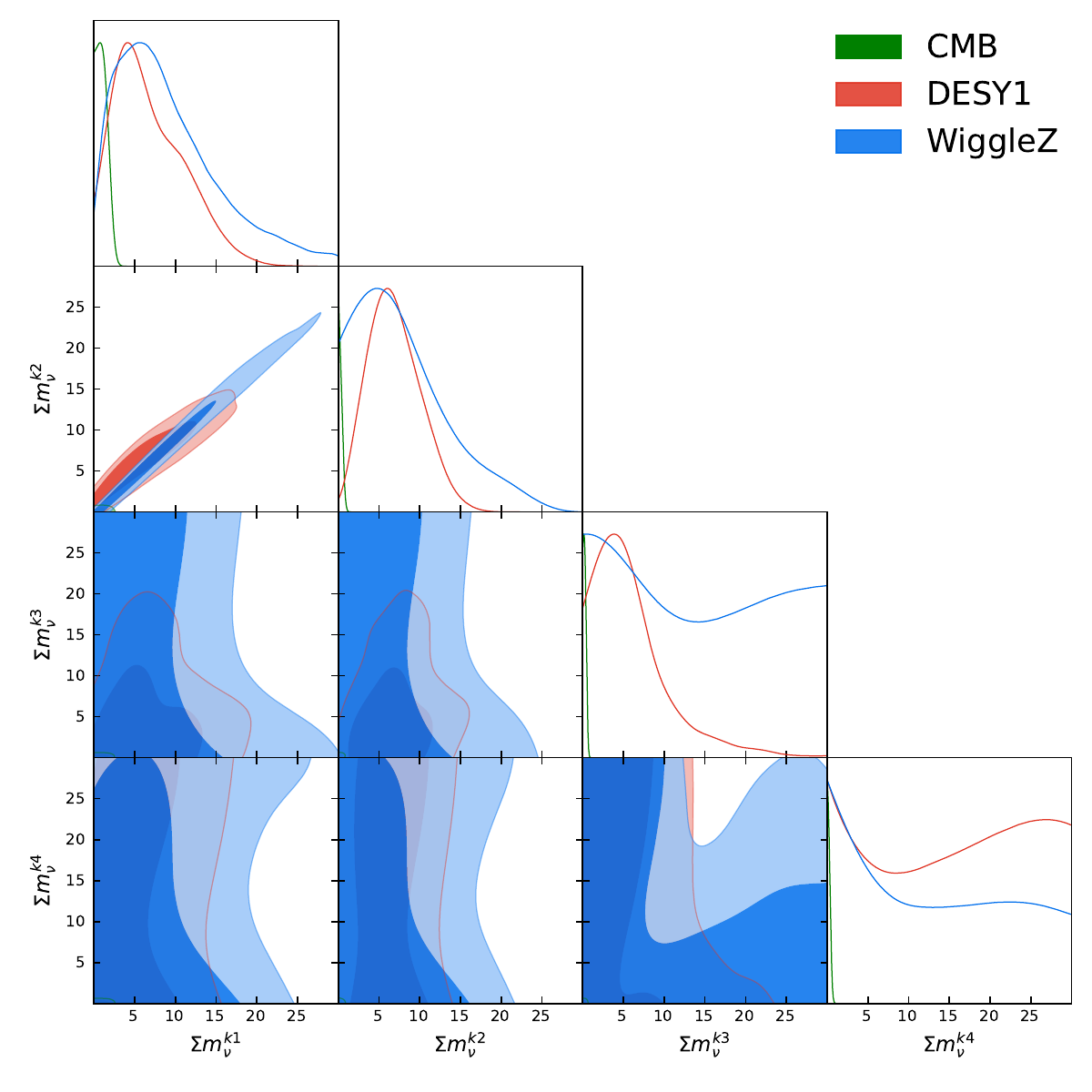}
	\caption{One-dimensional and two-dimensional posterior distributions of the neutrino mass parameters from different large scale structure probes in the $\sum m_\nu(k)$ model.}\label{fig:mnukindependent}
\end{figure*}

\begin{figure*}[htbp]
	\centering
	\includegraphics[scale=0.5]{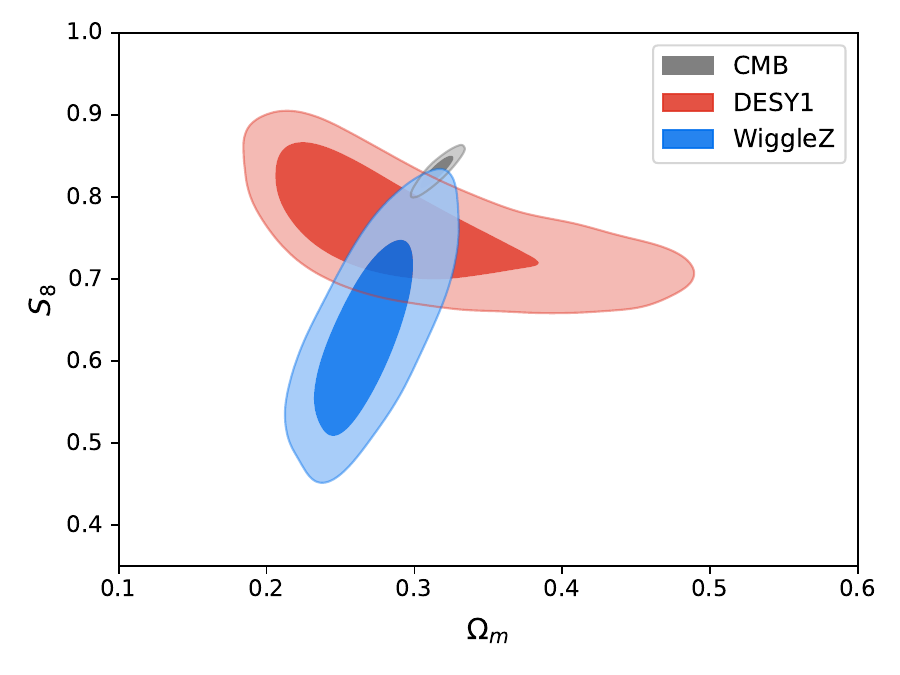}
	\includegraphics[scale=0.5]{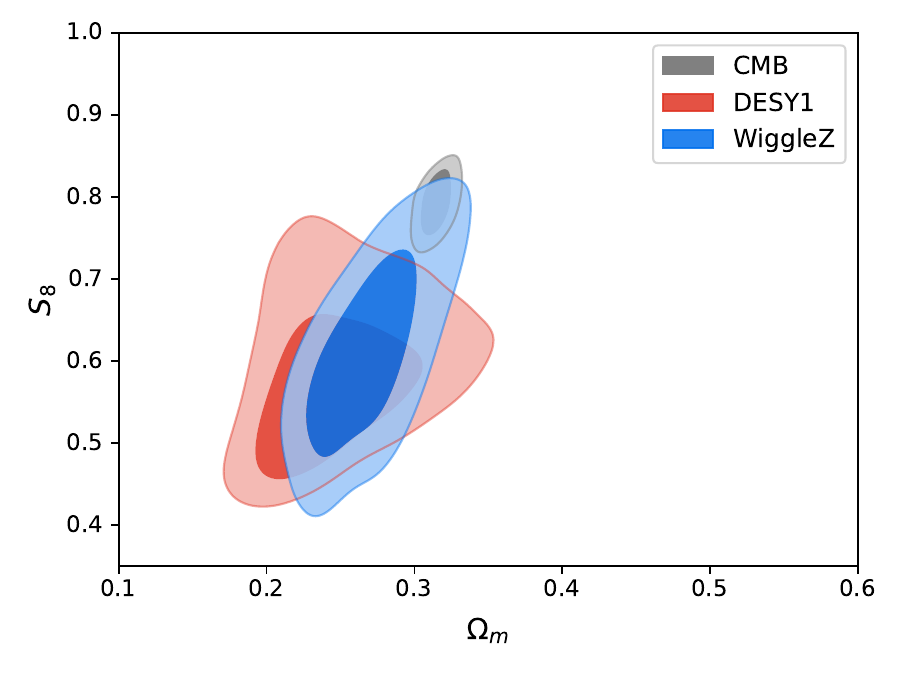}
	\caption{Two-dimensional posterior distributions of the parameter pair ($\Omega_m$, $S_8$) from different large scale structure probes in the $\Lambda$CDM ({\it left}) and $\sum m_\nu(k)$ ({\it right}) models, respectively.}\label{fig:S8}
\end{figure*}

\begin{figure*}[htbp]
	\centering
	\includegraphics[scale=0.5]{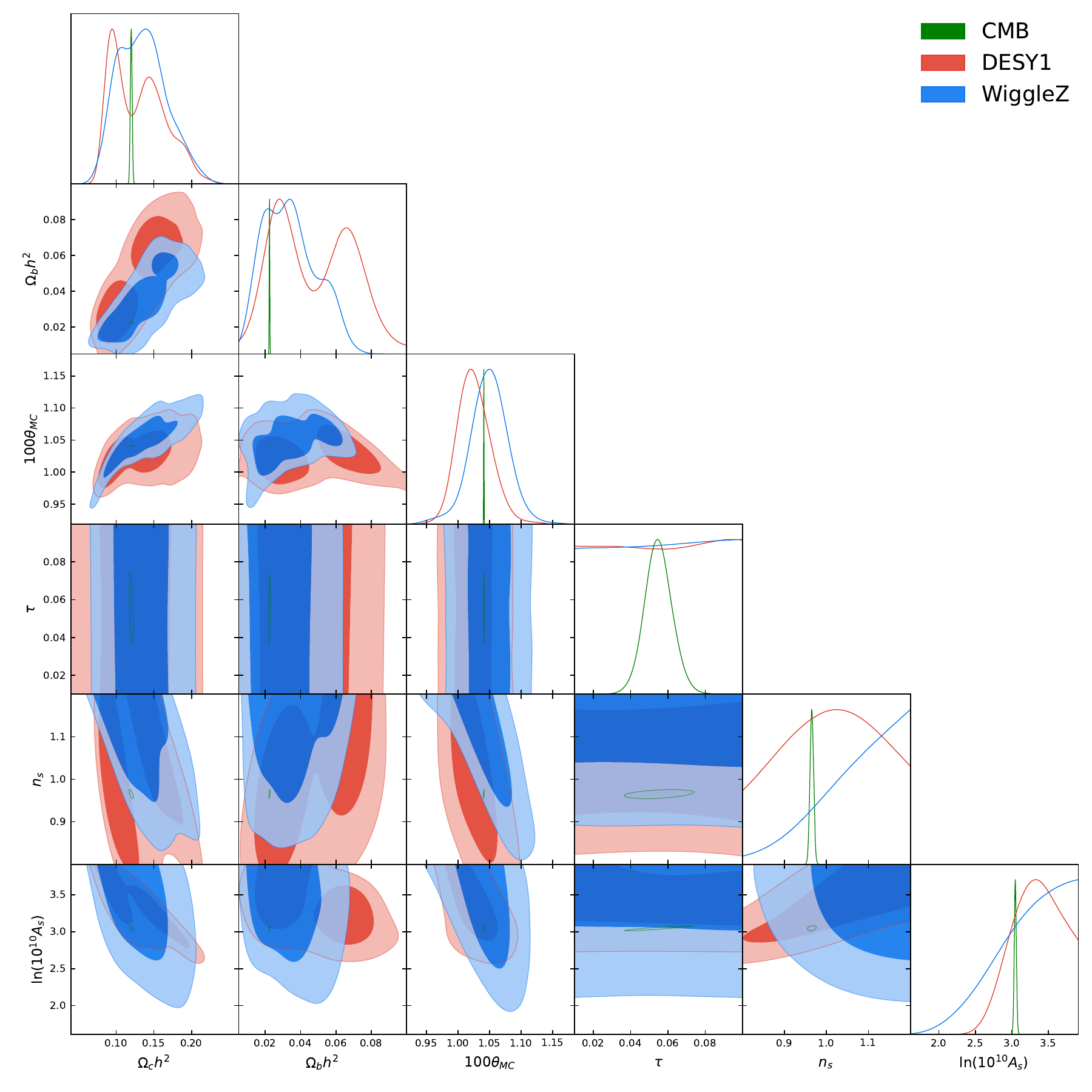}
	\caption{One-dimensional and two-dimensional posterior distributions of six cosmological parameters from different large scale structure probes in the $\sum m_\nu(k)$ model.}\label{fig:mnukindependent6params}
\end{figure*}

\begin{figure*}[htbp]
	\centering
	\includegraphics[scale=0.5]{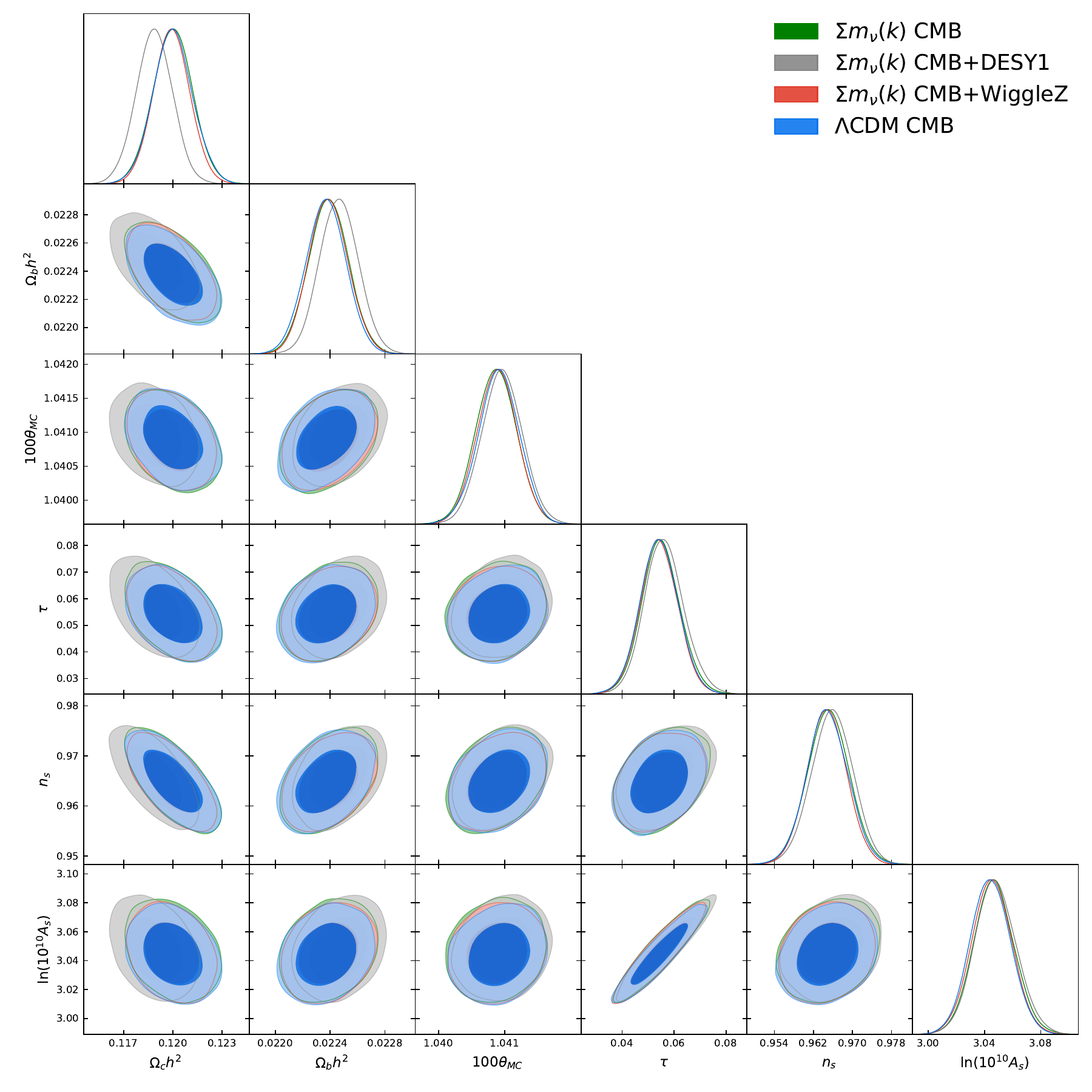}
	\caption{One-dimensional and two-dimensional posterior distributions of six cosmological parameters from different large scale structure probes in the $\sum m_\nu(k)$ model. As a comparison, we show the standard CMB-only constraint on $\Lambda$CDM. }\label{fig:mnuk4bins}
\end{figure*}

Constraining $\sum m_\nu$ in the $\Lambda$CDM+$\sum m_\nu$ model, we demonstrate that constraints on the globally single parameter $\sum m_\nu$ from various datasets are well consistent at $1\,\sigma$ confidence level (see Fig.\ref{fig:mnu}). The constraint from CW is weaker than that from CD when considering only one neutrino mass parameter $\sum m_\nu$. Interestingly, CBSDW gives a $2\,\sigma$ upper limit of $\sum m_\nu<0.121$ eV, which is weaker than $\sum m_\nu<0.096$ eV from CBS. This means that the addition of DESY1 \cite{DES:2017qwj1,DES:2017myr1,DES:2021wwk1} and WiggleZ \cite{Blake:2010xz1,Parkinson:2012vd1} weakens the constraint on $\sum m_\nu$. It seems unusual at first glance, but in reality, it is not. Since both DESY1 and WiggleZ prefer a value of $\sum m_\nu \sim 0.6$ eV, they pull $\sum m_\nu$ towards a slightly larger value in the combined constraint CBSDW, even if CBS has constrained $\sum m_\nu$ to a very small range. This also indicates that the statistical weight of DESY1 and WiggleZ in data analyses is non-negligible. Notice that here we do not use ACT DR6 lensing observations \cite{ACT:2023kun1} and consequently the CBS bound $\sum m_\nu<0.096$ eV is lower than $\sum m_\nu<0.072$ eV reported by the DESI collaboration \cite{DESI:2024uvr1,DESI:2024lzq1,DESI:2024mwx1}. 

\subsection{B. Constraints on $\sum m_\nu(z,\,k)$}

The CMB is a powerful tool for constraining a wide range of physical and cosmological parameters. Analyzing the temperature and polarization anisotropies in the CMB can help probe the fundamental laws of physics, the early universe physics, the nature of dark matter and dark energy and the properties of neutrinos. Up to now, only Planck CMB observations \cite{Planck:2019nip1,Planck:2018vyg1} allow us to implement a full redshift and scale search for massive neutrinos. The corresponding constraining results are shown in Fig.\ref{fig:mnuzk} and Tab.\ref{tab:mnuzk}. It is interesting that we find $\sum m_\nu^{52}=0.63^{+0.20}_{-0.24}$ eV, which is a $\sim 3\,\sigma$ evidence of neutrino mass at high redshift at intermediate scales. However, unfortunately, current CMB precision does not support us to perform a sensitive search in many other bins in the $\sum m_\nu(z,\,k)$ model. 

\subsection{C. Constraints on $\sum m_\nu(k)$}

Assuming the neutrino mass as a piecewise function of scale, we constrain the 4-bin $\sum m_\nu(k)$ model with three large scale surveys, i.e., Planck, DESY1 and WiggleZ. Overall, we find the tightest constraints sources from the largest scales, namely $k \in [0, 10^{-3}]$ h\,Mpc$^{-1}$ for all the data combinations. Very interestingly, the combination of CMB and WiggleZ gives $m_\nu^{k1}=0.75^{+0.20(1\sigma)+0.48(2\sigma)+0.67(3\sigma)}_{-0.27(1\sigma)-0.44(2\sigma)-0.52(3\sigma)}$ eV, indicating a beyond $5\,\sigma$ signal of $\sum m_\nu>0$ eV in $k \in [10^{-1}, +\infty)$ h\,Mpc$^{-1}$ (see Fig.\ref{fig:mnuk} and Tab.\ref{tab:mnuk}). Furthermore, the combination of CMB and DESY1 gives $\Sigma m_\nu^{k2}=0.55^{+0.27(1\sigma)+0.45(2\sigma)}_{-0.27(1\sigma)-0.54(2\sigma)}$ eV, which is also a $2\,\sigma$ evidence at intermediate scales in $k \in [10^{-2}, 10^{-1}]$ h\,Mpc$^{-1}$. From the viewpoint of observations, our results demonstrates that there truly exist massive neutrinos at small scales.

\subsection{D. Constraints on $\sum m_\nu(z)$}

Assuming the neutrino mass as a piecewise function of redshift, we constrain the 6-bin $\sum m_\nu(z)$ model with background and large scale structure observations from Planck, DESI, Pantheon+ \cite{Scolnic:2021amr1}, DESY1 and WiggleZ. Overall, we find that the tight constraints occur at high redshifts. Especially, the most constraining bounds in all the data combinations source from $\Sigma m_\nu^{4}$ in $z \in [10, 100]$ (see also Fig.\ref{fig:mnuz} and Tab.\ref{tab:mnuz}). CBSW gives the strongest constraint $\Sigma m_\nu^{4}<0.215$ eV. It is interesting that CBS and CBSW give $\Sigma m_\nu^{1}=1.01^{+0.47(1\sigma)+0.89(2\sigma)}_{-0.58(1\sigma)-1.00(2\sigma)}$ eV and $\Sigma m_\nu^{1}=0.65^{+0.25(1\sigma)+0.51(2\sigma)+0.67(3\sigma)}_{-0.25(1\sigma)-0.49(2\sigma)-0.60(3\sigma)}$ eV, which imply $2\,\sigma$ and $3\,\sigma$ evidences of massive neutrinos in $z \in [0, 1]$, respectively. Our results indicate that: (i) Background data can help determine the absolute neutrino mass scales; (ii) the constraints on $\Sigma m_\nu^{1}$ from CBS and CBSW is consistent within $1\,\sigma$ confidence level; (iii) DESI BAO measurements can strongly compress the neutrino mass parameter space in $z \in [1, 100]$.

\subsection{E. Evolutionary mechanisms of neutrino masses}
Since we have found five evidences of massive neutrinos including two $2\,\sigma$, two $3\,\sigma$ and one beyond $5\,\sigma$ in $\sum m_\nu(z,\,k)$, $\sum m_\nu(k)$ and $\sum m_\nu(z)$ models, it is natural and necessary to give possible evolutionary mechanisms of neutrino masses through the cosmic history. We propose four possible evolutionary mechanisms of neutrino masses over redshifts: (i) constant; (ii) freedom-suppression-freedom; (iii) freedom-suppression; (iv) suppression-freedom. After synthesizing all the information from our data analyses, we prefer the freedom-suppression-freedom scenario (see Fig.\ref{fig:mechanisms}), which is supported by the high-$z$ detection of $\Sigma m_\nu^{52}=0.63^{+0.20}_{-0.24}$ eV and low-$z$ evidences of $\Sigma m_\nu^{1}=1.01^{+0.47}_{-0.58}$ eV and $\Sigma m_\nu^{1}=0.65\pm 0.25$ eV. Actually, since $\Sigma m_\nu^{k1}=0.75^{+0.20}_{-0.27}$ eV and $\Sigma m_\nu^{k2}=0.55\pm 0.27$ eV are dominated by the low-$z$ large scale structure observations including WiggleZ and DESY1, they can also serve as low-$z$ evidences to support the freedom-suppression-freedom mechanism.

\subsection{F. Evidences for massive neutrinos independent of $S_8$ tension}
By constraining the 4-bin $\sum m_\nu(k)$ model with CMB, DESY1 and WiggleZ independently,  we have the following three points to demonstrate that the evidences reported here in the $\sum m_\nu(k)$ model are not induced by the $S_8$ tension. At first, in Fig.\ref{fig:mnukindependent}, one can easily find that both DESY1 and WiggleZ independently produce peaks of positive neutrino masses for $\sum m_\nu^{k1}$ and $\sum m_\nu^{k2}$. Especially, DESY1 gives $\sum m_\nu^{k2}=6.9^{+2.7(1\sigma)+6.3(2\sigma)+8.2(3\sigma)}_{-3.8(1\sigma)-5.9(2\sigma)-6.8(3\sigma)}$ eV indicating a $3\,\sigma$ evidence of neutrino mass, although this value is large due to the limited data precision.
This means both large scale structure probes independently prefer massive neutrinos in the absence of CMB observations. Furthermore, in Fig.\ref{fig:S8}, we find that our 4-bin $\sum m_\nu(k)$ model cannot resolve the $S_8$ tension and only slightly affect it. More specifically, the $S_8$ discrepancy is slightly weaker between CMB and WiggleZ, while slightly stronger between CMB and DESY1. This implies that massive neutrinos are not the origin of the $S_8$ anomaly.  
Finally, the signals of $\sum m_\nu^{52}>0$ eV in $\sum m_\nu(z,\,k)$ and $\sum m_\nu^{1}>0$ eV in $\sum m_\nu(z)$ sourcing from CMB and background probes (BAO and SN), respectively, can serve as important supporting evidences of massive neutrinos. After all, an undeniable fact is that these five evidences spanning high and low redshifts as well as small and intermediate scales are well compatible within $1\,\sigma$ confidence level. Overall, the addition of CMB to DESY1 or WiggleZ brings more accurate information of cosmological parameters (see also Figs.\ref{fig:mnukindependent6params} and \ref{fig:mnuk4bins}) and help determine the precise values of neutrino masses that DESY1 or WiggleZ prefers. As a consequence, similar to the strong evidences of dark energy in $\Lambda$CDM, the evidences of massive neutrinos also coexist with $S_8$ and $H_0$ tensions \cite{DiValentino:2020zio1}.

Very importantly, one cannot simply conclude from Fig.\ref{fig:mnukindependent} that the evidences of $\Sigma m_\nu^{k1}=0.75^{+0.20}_{-0.27}$ eV and $\Sigma m_\nu^{k2}=0.55\pm 0.27$ eV from CW and CD in the $\sum m_\nu(k)$ model, respectively, are produced by the inconsistencies of constraints on $\Sigma m_\nu^{k1}$ and $\Sigma m_\nu^{k2}$ between CMB and DESY1 (or WiggleZ). Actually, CMB alone cannot give good constraints on $\Sigma m_\nu^{k1}$ and $\Sigma m_\nu^{k2}$ at low redshifts at small scales (see $\sum m_\nu^{11}$, $\sum m_\nu^{12}$, $\sum m_\nu^{21}$ and $\sum m_\nu^{22}$ in Tab.\ref{tab:mnuzk}). The tight constraints are originated from the BFELF effect. For example, $\Sigma m_\nu^{62}<0.84$ eV dominates the constraining power in six $(z,\,k)$ bins (see $\sum m_\nu^{12}$, $\sum m_\nu^{22}$, $\sum m_\nu^{32}$, $\sum m_\nu^{42}$, $\sum m_\nu^{52}$ and $\sum m_\nu^{62}$ in Tab.\ref{tab:mnuzk}) in $k \in [10^{-2}, 10^{-1}]$ h\,Mpc$^{-1}$. This means that the inconsistencies source from high-$z$ CMB constraints and are not related to low-$z$ DESY1 and WiggleZ observations. Therefore, CMB, DESY1 and WiggleZ independently give consistent constraints on neutrino masses at low redshifts at small scales. 

In Fig.\ref{fig:mnukindependent6params}, it is easy to see that DESY1 and WiggleZ alone cannot provide accurate values for fundamental cosmological parameters. However, when combined with CMB, they produce good enough cosmologies, which are very close to the Planck $\Lambda$CDM cosmology \cite{Planck:2018vyg1}. Hence, the evidences of massive neutrinos reported here can be well approximately obtained under the Planck $\Lambda$CDM cosmology \cite{Planck:2018vyg1}. Additionally, the evidences of $\Sigma m_\nu^{k1}>0$ eV and $\Sigma m_\nu^{k2}>0$ eV from CW and CD can be easily understood in the following way. CMB brings the global cosmology information and high-$z$ information of massive neutrinos to calibrate the late-time large scale structure observations (DESY1 and WiggleZ) and break the parameter degeneracies, and conversely, large scale structure data brings the low-$z$ information of massive neutrinos to CMB and compensate for the lack of low-$z$ small-scale information in the CMB. The evidences of massive neutrinos are mainly originated from low-$z$ DESY1 and WiggleZ observations in the $\sum m_\nu(k)$ model.

\end{document}